\documentclass[aps,prd,nofootinbib,nobibnotes,twocolumn,floatfix,superscriptaddress,
letterpaper,showpacs,preprintnumbers]{revtex4-1}
\pdfoutput=1
\usepackage{graphicx}\usepackage{natbib}\usepackage{amsmath}
\usepackage{times}
\usepackage[T1]{fontenc}\usepackage{textcomp}\usepackage{mathptmx}
\usepackage{multirow}

\newcommand{\ba}{\begin{eqnarray}}
\newcommand{\ea}{\end{eqnarray}}
\newcommand{\no}{\nonumber}

\begin{document}
\title[]{A Flavor Sector for the Composite Higgs}

\author{Luca Vecchi}
\email{vecchi@lanl.gov}
\affiliation{Los Alamos National Laboratory, Theory Division T-2, Los Alamos, NM 87545, USA}

\begin{abstract}
We discuss flavor violation in large N Composite Higgs models. We focus on scenarios in which the masses of the standard model fermions are controlled by hierarchical mixing parameters, as in models of Partial Compositeness. We argue that a separation of scales between flavor and Higgs dynamics can be employed to parametrically suppress dipole and penguin operators, and thus effectively remove the experimental constraints arising from the lepton sector and the neutron EDM. The dominant source of flavor violation beyond the standard model is therefore controlled by 4-fermion operators, whose Wilson coefficients can be made compatible with data provided the Higgs dynamics approaches a ``walking" regime in the IR. Models consistent with all flavor and electroweak data can be obtained with a new physics scale within the reach of the LHC. Explicit scenarios may be realized in a 5D framework, the new key ingredient being the introduction of flavor branes where the wave functions of the bulk fermions end. 
\end{abstract}
\date{\today}
\preprint{LA-UR-12-22220}

\maketitle

\section{Introduction}
\label{intro}

Flavor violation in Composite Higgs (CH) models~\cite{KaplanGeorgi} was originally implemented via Extended Ultracolor interactions~\cite{ETC},~\footnote{Following~\cite{KaplanGeorgi} we will refer to the theory of the fundamental constituents  of the CH as Ultracolor, to distinguish it from Technicolor theories where the electroweak symmetry is not broken by a Higgs doublet.} and only more recently via the paradigm of Partial Compositeness~\cite{Contino:2003ve}. 

The Extended Ultracolor picture postulates the existence of a \emph{flavor sector} that mediates flavor-violating interactions between Standard Model (SM) and Ultracolor fermions at some high scale $m_F$. The effective field theory, renormalized at the scale $m_\rho\leq m_F$ where Ultracolor confines and the Higgs doublet $H$ is formed, contains operators of the form
\ba\label{1}
{\cal L}_{\rm EUC}&=&\alpha\left(\frac{m_\rho}{m_F}\right)^{\Delta-1}\overline{f}{H}f+\beta\, m_\rho^2\left(\frac{m_\rho}{m_F}\right)^{\Delta_{4\psi}-4}{H}^\dagger {H}\\\no
&+&\frac{\gamma}{m_F^2}\overline{f}f\overline{f}f+\frac{\delta}{m_F^2}\left(\frac{m_\rho}{m_F}\right)^{\Delta-1}\overline{f}\sigma_{\mu\nu}F^{\mu\nu}{H}f+\dots
\ea
Here $f$ collectively refers to the SM fermion fields, whereas $\alpha,\beta,\gamma,\delta$ are matrices in flavor space. In~(\ref{1}) we included the renormalization group effects induced by the strong Higgs dynamics and assumed that Ultracolor is in an approximate conformal phase below $m_F$, with $\Delta$ and $\Delta_{4\psi}$ denoting the scaling dimensions of the fields interpolating the Higgs and Higgs mass operators, respectively. 

The picture~(\ref{1}) has very general applicability, but does not tell us anything about the actual flavor structure of the couplings $\alpha,\beta,\gamma,\delta$ unless some assumption on the short distance physics is made. Partial Compositeness provides an efficient organizing principle for these couplings. 

In models of Partial Compositeness one assumes the SM fermions couple \emph{linearly} to composite operators of the Ultracolor sector via flavor-violating, anarchic couplings.~\cite{Kaplan} If the Higgs dynamics is nearly conformal for a large range of energies above $m_\rho$, then the mixing parameters can naturally be made hierarchical at low energies. The resulting effective field theory may still be formally written as in~(\ref{1}), but now with two important differences. First, in this model the flavor sector responsible for generating~(\ref{1}) is Ultracolor itself, so that the scale $m_F$ coincides with $m_\rho$ and the renormalization group factors in~(\ref{1}) disappear. Second, the matrices $\alpha,\beta,\gamma,\delta$ are \emph{hierarhical}, and can be employed to explain the SM fermion masses (see~\cite{Grossman:1999ra}\cite{Gherghetta:2000qt} for an implementation of this idea within 5D dual scenarios).

The assumption $m_F=m_\rho$ has the very attractive feature of relating flavor violation to the weak scale, and thus leading to a very predictive scenario. However, the stringent experimental bounds on the Wilson coefficients of the \emph{dipole operators} -- especially from $\mu\to e\gamma$, the electron EDM, and the neutron EDM -- severely constrain such program~\cite{APS}\cite{Agashe:2006iy}\cite{Davidson:2007si}\cite{noi}.

The aim of this paper is to show that these constraints can be significantly alleviated by considering scenarios in which both SM \emph{and} Ultracolor fermions are partially composite states of a strongly coupled, nearly conformal flavor sector characterized by a large dynamical mass $m_F$. In these scenarios the SM flavor problem can be addressed at a higher scale $m_F>m_\rho$, and the dangerous dipole operators in~(\ref{1}) suppressed by a renormalization group factor of order $m_\rho^2/m_F^2$ compared to conventional models of Partial Compositeness.

In Section~\ref{general} we present the model in detail. We then identify a class of realistic scenarios in Section~\ref{realistic}, and discuss their dual 5D picture in Section~\ref{5D}. We finally conclude in Section~\ref{discussion}.

\section{The Model}
\label{general}

The main building blocks of our model are the SM fermion and gauge fields, a Higgs sector, and a \emph{flavor sector}. 

The Higgs sector is basically a copy of the one postulated in CH models~\cite{KaplanGeorgi}, and consists of a strongly coupled Ultracolor theory that after chiral symmetry breaking delivers a Nambu-Goldstone boson (NGB) multiplet containing a field with the quantum numbers of the Higgs doublet.

The existence of a separate flavor sector is the main novelty of our approach to Partial Compositeness. The flavor sector is a strongly coupled and approximately conformal field theory. Its couplings to the SM fermions control the only source of flavor violation in the model. 

The quantum numbers of the Ultracolor and flavor sectors are chosen in such a way that the SM and Ultracolor fermions dominantly communicate with each other via the weak gauging of the SM group and the nearly conformal flavor theory. At some high UV cutoff scale $\Lambda$ the SM fermions $f$ and the ultra-fermions $\psi$ are assumed to couple linearly to composite operators $O_{f,\psi}$ of the flavor sector:
\ba\label{mixing}
\lambda_f O_ff+\lambda_\psi O_\psi \psi+{\rm h.c.},
\ea
with $\lambda_{f,\psi}$ dimensionless parameters. No direct coupling between the $f$s and the $\psi$s is present, up to irrelevant terms suppressed by inverse powers of the cutoff. These assumptions should be compared to those of the Partial Compositeness scenario of~\cite{Kaplan}, where the operators $O_{f,\psi}$ themselves are composites of the $\psi$s.

At much smaller scales the approximate conformal invariance of the flavor sector is broken and a dynamical mass of order $m_F\ll\Lambda$ is generated. We will assume that this phenomenon is not accompanied by the formation of light exotic states (see however the end of this section).

According to Naive Dimensional Analysis (NDA)~\cite{NDA}, the low energy theory renormalized at the scale $m_F$ contains (up to corrections suppressed by powers of $m_F/\Lambda$) the SM and Ultracolor Lagrangians plus
\ba\label{NDA}
{\cal L}_{\rm EUC}&=&\frac{m_F^4}{g_F^2}\,\widehat{\cal L}\left(\frac{\epsilon_fg_Ff}{m_F^{3/2}}, \frac{\epsilon_\psi g_F\psi}{m_F^{3/2}},\frac{D_\mu}{m_F}\right)+\cdots,
\ea
where $\widehat{\cal L}$ is a functions with order one entries and $D_\mu$ a covariant derivative. Here $g_F$ is the coupling characterizing the strength of the flavor sector, and by NDA it satisfies 
\ba
g_F\lesssim4\pi, 
\ea
with $g_F\ll4\pi$ only in theories with a large number of fundamental constituents. Possible loop corrections of order $g_F^2/16\pi^2$ are included in the dots of~(\ref{NDA}). 

The assumption that the theory can be characterized by a single coupling and a single scale is very restrictive, but should suffice to provide a quantitative understanding of the model. 

The coefficients
\ba
\epsilon_{f,\psi}\sim\frac{\lambda_{f,\psi}(m_F)}{g_F}\lesssim1
\ea
measure the amount of compositeness of the associated field, such that for $\epsilon_{f,\psi}\sim1$ the corresponding fermion behaves like a bound state of the strong flavor dynamics. The magnitude of the $\epsilon_{f,\psi}$s is controlled by the RG evolution of the $\lambda_{f,\psi}$s from the cutoff scale down to $m_F\ll\Lambda$, and can naturally be hierarchical. The existence of hierarchical mixing parameters is at the heart of the Partial Compositeness solution to the SM flavor puzzle.

The Lagrangian~(\ref{NDA}) describes the dominant interactions (beyond those induced by the SM gauge group) between the Higgs sector fields $\psi$ and the SM fermions $f$. The formal structure of ${\cal L}_{\rm EUC}$ is analogous to the one in Extended Ultracolor models, see Eq.(\ref{1}). The crucial new feature is the presence of the \emph{hierarchical} flavor-violating parameters $\epsilon_{f,\psi}$. 

The rest of the paper will be devoted to a detailed analysis of the effective field theory~(\ref{NDA}). While we derived~(\ref{NDA}) from the principles of Partial Compositeness, our results will be more general and will in fact apply to any theory of flavor leading to the form~(\ref{NDA}), such as for example Froggatt-Nielsen models.

The scenarios of Partial Compositeness studied in the previous literature can be recovered from~(\ref{NDA}) in the limit in which the mass scales and couplings of the flavor and Ultracolor sectors coincide, and $\epsilon_\psi\to1$.

\subsection{Yukawa Interactions}

The ultra-fermion bilinear $\overline{\psi}\psi$ is assumed to be the operator with the lowest scaling dimension that interpolates the composite Higgs field. Using~(\ref{NDA}), we thus see that in our model the Yukawa matrices for the charged SM fields are effectively described by the following operators:
\ba\label{Yukawa}
\frac{g_F^2}{m_F^2}\epsilon_\psi^2\,\epsilon_{f^a_i}\epsilon_{f^b_j}(\overline{\psi}\psi)_{m_F}(\overline{f^a_i}f^b_j)+{\rm h.c.},
\ea
with $f^a_i=q_i, u_i, d_i, \ell_i, e_i$ the five SM fermion representations and $i,j=1,2,3$ family indices. Note that the above relation constrains the spurionic symmetries carried by the $\lambda_{f,\psi}$s. (The Yukawa operator~(\ref{Yukawa}) should be compared to the one found in the Partial Compositeness model of~\cite{Kaplan}, where multiple insertions of the ultra-fermion bilinear typically appear.)

Non-hierarchical neutrino masses and mixing may be generated as discussed in~\cite{noi}. In this paper we will only be interested in the physics of the charged leptons and quarks, for which much stronger experimental constraints exist.

Including the RG flow down to the scale $m_\rho$ at which the Ultracolor theory becomes strong and the Higgs doublet $H$ emerges, and assuming approximate conformal invariance in the mass range $m_\rho<\mu<m_F$, we find that the Yukawa matrices arising from~(\ref{Yukawa}) have the structure anticipated in Eq.~(\ref{1}), namely
\ba\label{y}
y_{ij}\sim\frac{g_F^2}{g_\rho}\epsilon_\psi^2\,\epsilon_{f^a_i}\epsilon_{f^b_j}\left(\frac{m_\rho}{m_F}\right)^{\Delta-1}.
\ea
$\Delta$ is the scaling dimension of the ultra-fermion bilinear, and $g_\rho$ is the typical coupling among the resonances of the Higgs sector. Again, NDA suggests that $g_\rho\sim4\pi$ in maximally strong theories, while $g_\rho<4\pi$ in large $N$ Ultracolor. In deriving~(\ref{y}) from~(\ref{Yukawa}) we adopted the NDA estimate $(\overline{\psi}{\psi})_{m_\rho}\sim m_\rho^2 H/g_\rho$.

As in the standard scenarios of Partial Compositeness, the Yukawa matrix~(\ref{y}) can elegantly generate the SM flavor hierarchy. The new feature of our model is the existence of a universal suppression $(m_\rho/m_F)^{\Delta-1}$, which can be traced back to the irrelevance of the Yukawa interaction~(\ref{Yukawa}). Because of this suppression, it is generally {\it impossible} to decouple the flavor sector while keeping the SM fermion masses fixed. This is especially true in theories in which $\Delta\gg1$, where the renormalization group suppression is significant, but may be alleviated if $\Delta$ is not far from one. 

For a scaling dimension too close to one the Higgs will behave like a fundamental field, its mass operator will become strongly relevant, and one will end up with the naturalness problem typical of weakly coupled Higgs sectors. This may not occur in small $N$ theories, where it is conceivable that $\Delta\sim1$ be compatible with $\Delta_{4\psi}\gtrsim4$~\cite{CTC}. Unfortunately, the strongly coupled nature of these theories makes it difficult to establish whether such possibility can actually be realized within simple 4D theories. 

Large $N$ field theories have the great advantage of having a small expansion parameter, and therefore being more tractable and predictive. In this class of models one finds $\Delta_{4\psi}\simeq2\Delta$, in which case a solution of the hierarchy problem requires 
\ba\label{2}
\Delta\gtrsim2. 
\ea
Arguments based on the large $N$ and/or the ladder expansions indicate that asymptotically free, non-supersymmetric, non-abelian theories indeed satisfy Eq.~(\ref{2}), with $\Delta$ approaching 2 at the lower end of the conformal window (see~\cite{Cohen:1988sq} for a pedagogical presentation and for earlier references). This claim has been recently substantiated by lattice simulations~\cite{Appelquist:2012nz}.

In this paper we focus on natural, large $N$ CH models and assume that Ultracolor enters a strongly coupled and nearly conformal phase satisfying~(\ref{2}) below a scale $\Lambda_W$ not far from $m_F$.~\footnote{Because in our model~(\ref{mixing}) the flavor sector is also assumed to carry Ultracolor, the condition $\Lambda_W\lesssim m_F$ might seem to be necessary in order to safely view the Ultracolor group as a ``weak gauging" of part of the global symmetry of the flavor sector, and in particular justify Eq.~(\ref{NDA}). We emphasize, however, that given the order one uncertainty already included in~(\ref{NDA}), we do not see any obvious inconsistency in taking $\Lambda_W$ slightly above $m_F$. Heuristic arguments suggest that asymptotically free theories enter a strongly coupled conformal window (``walking" Ultracolor) when the gauge coupling is $\sim\sqrt{4\pi}$, and hence small compared to the value $g_F\sim4\pi$ we expect at the matching scale $m_F$. It will soon become clear that the most favorable condition is in fact $\Lambda_W\gtrsim m_F$, so we will stick to that case in what follows. We anticipate that the corresponding uncertainty encountered in~(\ref{NDA}) is at most of order $1/\sqrt{4\pi}<O(1)$, and hence irrelevant. 

If the reader feels uncomfortable with the above reasoning, we emphasize that it is possible to consider different realizations, in which the mixing $O_\psi\psi$ is replaced by $O_\Psi\Psi(\psi)$, where now $\Psi(\psi)$ is an Ultracolor-singlet composite of $\psi$s. This latter model is clearly consistent with the assumption $\Lambda_W>m_F$, and would lead to an analogous low energy phenomenology.} This assumption introduces perhaps the main source of uncertainty in our discussion, because no fully reliable tool able to determine the relevance of the operators of a strong dynamics is currently known. In cases where we are not able to unambiguously identify the most relevant operator we will thus be forced to rely on heuristic arguments and sometimes even assumptions. The main conclusions of this paper should hold, if not generally, at least in a class of realistic strongly coupled theories.

We will now discuss the impact of the scale $m_F$ on flavor violation in more detail.

\subsection{Dimension-6 Operators}
\label{dim6}

In order to make a comparison with the standard scenarios of Partial Compositeness (see~\cite{noi} for a recent discussion) it is convenient to introduce the following notation:
\ba\label{yy}
\tilde\epsilon_{f}\equiv \epsilon_{f}\frac{g_F}{g_\rho}\epsilon_\psi\left(\frac{m_\rho}{m_F}\right)^{(\Delta-1)/2}.
\ea
The $\tilde\epsilon_f$s formally measure the amount of mixing between the SM fermions and the resonances of the Higgs dynamics, and should be compared to the $\epsilon_f$s that instead control the mixing with the flavor sector. With this definition the Yukawa coupling~(\ref{y}) becomes $y_{ij}\sim g_\rho \tilde\epsilon_{f^a_i}\tilde\epsilon_{f^b_j}$, and hence is formally the same as that found in models where the Higgs and flavor sectors coincide. A fair comparison between the standard framework and our scenario can thus be given if $g_\rho$, $m_\rho$ as well as the $\tilde\epsilon_f$s are taken to be the same in both theories. This way we are left with two new parameters: $m_\rho/m_F$ and $g_\rho/g_F$.

Flavor violation beyond the renormalizable level is controlled by 4-fermion interactions, as well as dipole and penguin operators. From the NDA Lagrangian~(\ref{NDA}), and using the definition~(\ref{yy}), the coupling of a typical 4-SM fermion operator is easily found to be of the form
\ba\label{4f}
&&\epsilon_{f^a_i}\epsilon_{f^b_j}\epsilon_{f^c_k}\epsilon_{f^d_l}\frac{g_F^2}{m_F^2}\\\no
&=&\tilde\epsilon_{f^a_i}\tilde\epsilon_{f^b_j}\tilde\epsilon_{f^c_k}\tilde\epsilon_{f^d_l}\frac{g_\rho^2}{m_\rho^2}\times\left(\frac{1}{\epsilon_\psi}\right)^4\left(\frac{g_\rho}{g_F}\right)^2\left(\frac{m_\rho}{m_F}\right)^{4-2\Delta}.
\ea
The first factor on the right hand side of~(\ref{4f}) represents the standard result in models of Partial Compositeness, while the last characterizes our scenario. This latter factor may be understood by noting that the coefficient of the 4-$f$ operator roughly goes as $y^2(m_F)/m_F^2\sim y^2(m_\rho)(m_F/m_\rho)^{2\Delta-2}/m_F^2$.

In order to minimize the effect of the 4-$f$ operators we require
\ba\label{epspsi}
\epsilon_\psi\simeq1, 
\ea
and interpret Ultracolor as the low energy remnant of the flavor dynamics. From this perspective the $\psi$s are not coupled linearly to the flavor sector as shown in~(\ref{mixing}), but rather arise as composite states at the scale $m_F$. The nearly conformal flavor sector and ``walking" Ultracolor are now different phases of a single flavorful, strong dynamics. This alternative view leads to the very same low energy phenomenology as the one following from~(\ref{mixing}) and~(\ref{epspsi}).

We will assume that the condition~(\ref{epspsi}) holds throughout the paper. With this condition and our working hypothesis~(\ref{2}) we see that a suppression of the 4-fermion operators is possible for
\ba\label{sup4f}
\Delta\simeq2,~~~~~~~~~~~~{\rm and}~~~~~~~~~~~g_\rho\lesssim g_F.
\ea
While the condition $\Delta\simeq2$ is typical of the ``walking" regime of asymptotically free non-abelian theories, the suppression $g_\rho^2/g_F^2<1$ can only be achieved in CH models with a moderately large number of fundamental constituents and a relatively strong flavor dynamics.

In the case of the dipole operators the most relevant contribution arises from
\ba\label{dipole}
&&\epsilon_{f^a_i}\epsilon_{f^b_j}\epsilon_\psi^2\frac{g_F^2}{m_F^4}(\overline{f^a_i} \sigma_{\mu\nu} F^{\mu\nu} f^b_j)(\overline\psi\psi)_{m_F}\\\no
&\to&\tilde\epsilon_{f^a_i}\tilde\epsilon_{f^b_j}\frac{g_\rho}{m_\rho^2}(\overline{f^a_i} \sigma_{\mu\nu} F^{\mu\nu} f^b_j)H\times\frac{m_\rho^2}{m_F^2},
\ea
where in the second step we included the effect of the RG flow down to the low energy scale $m_\rho$. As anticipated in Section~\ref{intro}, here we find a generic suppression of order $m_\rho^2/m_F^2$ compared to the usual scenario,~\footnote{Dipole operators often come with a loop factor in theories where the quantum numbers and couplings of the low energy resonances are \emph{not} generic. In these models the suppression of the dipole operators we show in Eq.~(\ref{dipole}) is accompanied by a factor of $g_F^2/g_\rho^2$. We will focus on generic, strongly coupled theories in what follows, but the reader should keep in mind that explicit realizations of our scenario (for example in a 5D dual picture) having minimal field content may manifest a reduced suppression if $g_F>g_\rho$.\label{foot}} which originates from the fact that this operator has a scaling dimension larger than~(\ref{Yukawa}) by two units. Possible additional contributions from less relevant operators are by definition down by higher powers of $m_\rho/m_F$ and will be ignored.

The relevance of penguin operators is unfortunately more model-dependent, as it necessarily depends on the (uncalculable) anomalous dimensions of the associated $\psi$ composite operators. An unambiguous prediction can only be given for the operator $(\overline{f}\gamma^\mu f)(\overline\psi\gamma^\mu\psi)$, where $(\overline\psi\gamma^\mu\psi)$ denotes the conserved current associated to the SM $SU(2)\times U(1)$ gauge symmetry. By charge conservation this operator receives small RG corrections from the Ultracolor dynamics (at leading order in the SM gauge couplings its scaling dimension is exactly 3), and would provide the dominant contribution in a weakly coupled theory. Neglecting the RG flow down to the scale $m_\rho$ and making use of~(\ref{NDA}), (\ref{yy}), and NDA, we thus have:
\ba\label{penguin}
&&\frac{g_F^2}{m_F^2}\epsilon_{f^a_i}\epsilon_{f^b_j}\epsilon_\psi^2(\overline{f^a_i}\gamma^\mu f^b_j)(\overline\psi\gamma^\mu\psi)_{m_F}\\\no
&\to&\frac{g_\rho^2}{m_\rho^2}\tilde\epsilon_{f^a_i}\tilde\epsilon_{f^b_j}(\overline{f^a_i}\gamma^\mu f^b_j)H^\dagger i{D}_\mu H\times\left(\frac{m_\rho}{m_F}\right)^{3-\Delta}.
\ea
Again, the second factor is a feature of our scenario. Here we find that the penguin operators are suppressed when $\Delta<3$, a condition which is less stringent than~(\ref{sup4f}). Other operators may contribute to~(\ref{penguin}); having no control over them, we will assume that they are more irrelevant than the one shown above, so that our conclusions are unchanged.

In summary, we have shown that in models where the flavor and Higgs scales are not coincident, the dipole as well as penguin operators are generally suppressed by powers of $m_\rho/m_F$ compared to the standard scenarios of Partial Compositeness. Under our working hypothesis~(\ref{epspsi}) the 4-fermion operators may or may not be parametrically suppressed depending on whether Eq.~(\ref{sup4f}) is satisfied or not.

\section{A Class of Realistic Scenarios}
\label{realistic}

The most stringent constraints on the standard scenarios of Partial Compositeness arise from the neutron EDM~\cite{APS}, the electron EDM, as well as lepton flavor violation, in particular $\mu\to e\gamma$~\cite{Agashe:2006iy}\cite{Davidson:2007si} (see~\cite{noi} for a recent discussion of all the constraints). The electroweak precision bounds represent no serious obstacle provided $g_\rho v/m_\rho<1$.

In this section we will show that \emph{quark} flavor violation may be suppressed in the models of Section~\ref{general}, but not completely removed, such that the resulting theory will be rather predictive. On the other hand, we will argue that the presence of experimental signatures in the \emph{lepton} sector does not represent a generic expectation, since our framework offers a very simple solution to the lepton flavor problem of CH models. 

Because the flavor and Higgs sectors are characterized by two distinct dynamical masses, a priori there is no reason why the strong flavor sector responsible for generating quark flavor and the one generating lepton flavor should have the same confinement scale. In fact, an obvious possibility is that the flavor sector be split in two distinct strongly coupled theories with two very well separated dynamical scales. (For example, we can arrange one of the two theories to include color as a weakly gauged subgroup of the chiral symmetry, while the other to be only invariant under the weak force.) The Extended Ultracolor \emph{lepton} interactions in~(\ref{NDA}) may then be generated at a scale $m_{F_\ell}$ much higher than the mass scale at which the \emph{quark} flavor hierarchy is addressed, 
\ba
m_{F_q}\ll m_{F_\ell},
\ea 
realizing a mechanism somewhat reminiscent to the one invoked in ``tumbling Extended Ultracolor." We thus conclude that in these models quark flavor violation will in general be controlled by the parameter $m_\rho/m_{F_q}$, while lepton violation by $m_\rho/m_{F_\ell}\ll m_\rho/m_{F_q}$. This will allow us to effectively decouple lepton flavor violation, as we will argue shortly.

\subsection{Quark Sector}
\label{quarksector}

Let us first discuss in some detail flavor violation in the quark sector, and in particular the impact of the new parameters $g_\rho/g_{F_q}$ and $m_\rho/m_{F_q}$.

An upper bound on the scale $m_F$ is obtained by imposing $\epsilon_{f,\psi}\lesssim1$ in~(\ref{y}),
\ba\label{lower}
\left(\frac{m_\rho}{m_{F}}\right)^{\Delta-1}\gtrsim\frac{y_f}{g_F}\frac{g_\rho}{g_F},
\ea
and is saturated in the limit $\epsilon_{f,\psi}\simeq1$. It is understood that in the quark sector $y_f=y_t\sim1$. 

We see from~(\ref{lower}) that even with a generic parameter choice such as $g_\rho\sim g_{F_q}\sim4\pi$ a suppression of two order of magnitudes for the dipole operators~(\ref{dipole}) may be arranged in ``walking" Ultracolor theories. The situation significantly improves in moderately large $N$ CH models and maximally strong flavor sectors, where $g_\rho<g_{F_q}$ is possible. In that case not only is the flavor scale parametrically larger, but even the 4-$f$ operators will be suppressed. Such effects would easily render the model compatible with the very stringent bounds from the neutron EDM and Kaon oscillation if $m_\rho$ is in the few TeV range~\cite{noi}.

\subsection{Lepton Sector}
\label{leptonsector}

As opposed to the quark sector, a parametric enhancement of $m_{F_\ell}/m_\rho$ can be arranged without imposing any condition on $g_\rho/g_{F_\ell}$. This is so because experimentally we know that the mixing between the SM and the composites of the Higgs sector have to be small $\tilde\epsilon_{\ell,e}\ll1$, but this fact does not constrain the mixing with the flavor sector at the scale $m_{F_\ell}$. Now, if the $\epsilon_{\ell,e}$s are large, say of the same order as the $\epsilon_{q,u,d}$s, then to generate small Yukawa couplings for the leptons at the scale $m_\rho$ one is forced to take $m_\rho\ll m_{F_\ell}$. 

(Said differently, in this model the mass hierarchy among the three families is explained by hierarchical $\tilde\epsilon_f$s, while the hierarchy between quark and lepton masses may arise from the assumption $m_{F_\ell}\gg m_{F_q}$.)

To be more quantitative, we re-write the definition~(\ref{yy}) in a very instructive way: 
\ba\label{ass}
\frac{m_\rho}{m_F}\simeq\left(\frac{g_\rho}{g_F}\frac{\tilde\epsilon_{f}}{\epsilon_{f}}\right)^{\frac{2}{\Delta-1}},
\ea
where the symbol $\simeq$ follows from our assumption~(\ref{epspsi}).~\footnote{Physically, Eq.~(\ref{ass}) expresses the fact that the ratio between the couplings of the SM fermions to the Higgs sector (renormalized at $m_\rho$) and the couplings of the SM fermions to the flavor sector (renormalized at $m_F$) is controlled by the scale separation. The irrelevance of the Yukawa operator~(\ref{Yukawa}) implies that this ratio cannot be greater than one.} From~(\ref{ass}) it is apparent that a large scale separation can easily arise from the requirement $\tilde\epsilon_{\ell,e}\ll\epsilon_{\ell,e}$ if $\Delta$ is not too large.

The dominant experimental constraints here come from $\mu\to e\gamma$ and the electron EDM. Ref.~\cite{noi} shows that with $m_\rho\sim10$ TeV a suppression of the dipole operators on the order of $10^{-2}-10^{-3}$ would be needed to make the standard scenarios viable. From Eq.~(\ref{dipole}) we see that this suppression can be obtained taking $(m_\rho/m_{F_\ell})^2\sim10^{-2}-10^{-3}$, a condition which for $\Delta\sim2$ is very far from saturating the lower bound~(\ref{lower}) with $y_f=y_\tau\sim10^{-2}$. The lepton flavor problem emphasized in~\cite{Agashe:2006iy}\cite{Davidson:2007si}\cite{noi} is thus easily evaded.

As discussed in~\cite{Davidson:2007si} the effect of the 4-fermion operators with either 4 lepton fields or 2 quarks and 2 leptons is well below experimental sensitivity because of the small lepton parameters $\tilde\epsilon_{e,\ell}$s. Therefore, such operators do not imply any significant constraint on these models.

\subsection{The Higgs Mass}
\label{Higgssec}

The Higgs doublet is assumed to be an exact NGB of the Ultracolor sector, and to acquire a potential by the \emph{explicit} breaking of the coset symmetry. In general, the present construction predicts two sources of explicit symmetry breaking, arising from the couplings $\lambda_f$ and $\lambda_\psi$ in Eq.~(\ref{mixing}). However, the breaking mediated by $\lambda_\psi$, and effectively parametrized below $m_F$ by operators involving a few ultra-fermions, generally results in very large corrections to the Higgs mass because of our choice~(\ref{epspsi}). In order to avoid these unattractive effects it is sufficient to assume that the flavor sector realizes the global symmetry of the strong Higgs dynamics linearly, such that only the couplings $\lambda_f$s will contribute to the Higgs potential. 

The main source of coset symmetry breaking can be parametrized at the scale $m_\rho$ by the top (proto)Yukawa coupling and 4-$\psi$ operators, see the first two terms in~(\ref{1}). (Operators with higher powers of ultra-fermions are subleading in an expansion in powers of $g_F^2/g_\rho^2\times (m_\rho/m_F)^{\Delta_{(2n+2)\psi}-\Delta_{2n\psi}}$, where in the exponent we defined the difference between the scaling dimensions of the $(2n+2)$-$\psi$ and $2n$-$\psi$ operators.) 

Now, integrating out SM and Ultracolor fluctuations above $m_\rho$ at one-loop and at all orders, respectively, the most general 4-$\psi$ contribution is found to scale as
\ba\label{4psi}
\frac{g_F^2N_f}{16\pi^2}\epsilon_\psi^4\frac{g_F^2}{m_F^2}\sum_\alpha c_\alpha\,\epsilon_f^{2n_\alpha} (\overline\psi\psi\overline\psi\psi)^{(\alpha)}_{m_\rho}\left(\frac{m_\rho}{m_{F}}\right)^{\Delta^{(\alpha)}_{4\psi}-6}.
\ea
In the above expression $N_f$ is the number of SM fermions involved in the loop (for example, in the case of the top $N_f\to N_c=3$, the number of colors), $\Delta_{4\psi}^{(\alpha)}$ is the scaling dimension of the 4-ultra-fermion operator with structure $\alpha$ (renormalized at the scale $m_\rho$), $c_\alpha$ are order one coefficients, and finally $n_\alpha=1,2,\cdots$ denotes the power of $\epsilon_f$s required to construct an invariant. 

(We emphasize that the fermion with the largest $\epsilon_f$ is not necessarily the top quark: if $m_{F_\ell}$ is very large, the tau contribution might in fact dominate. Yet, for the typical values discussed in the previous subsections it is the top quark that leads to the dominant effects. We will assume this is the case here, as well.)

Using NDA, Eq.~(\ref{4psi}) leads to the following ``UV contribution" to the Higgs potential
\ba\label{UV}\no
\delta V_{\rm UV}&\sim&\frac{N_f}{16\pi^2}m_\rho^4\sum_\alpha c_\alpha\,\epsilon_f^{2n_\alpha}\; \widehat V^{(\alpha)}\left(\frac{g_\rho H}{m_\rho}\right)\times\frac{g_F^4}{g_\rho^4}\epsilon_\psi^4\left(\frac{m_\rho}{m_{F}}\right)^{\Delta^{(\alpha)}_{4\psi}-4}\\
&+&\cdots,
\ea
where the dots refer to the contributions from more irrelevant operators and higher loops, and the $\widehat V^{(\alpha)}$s are dimensionless functions with order one entries. The condition $\Delta^{(\alpha)}_{4\psi}>4$ expresses the absence of sensitivity to the UV cutoff. 

The potential~(\ref{UV}) reduces to the one found in the standard scenarios when $m_F\to m_\rho$ and $g_F\to g_\rho$. In this limit operators with an arbitrary number of $\psi$s should be considered on an equal footing as the 4-$\psi$ ones. 

The other irreducible contribution to the Higgs potential arises from top-quark virtualities \emph{below} $m_\rho$. This is of the form
\ba\label{IR}
\delta V_{\rm IR}&\sim&\frac{N_c}{16\pi^2}m_\rho^4\,\frac{y_t^2}{g_\rho^2}\,\widehat V\left(\frac{g_\rho H}{m_\rho}\right)\\\no
&\sim&\frac{N_c}{16\pi^2}m_\rho^4\,(\epsilon_{q_3}\epsilon_{u_3})^2\,\widehat V\left(\frac{g_\rho H}{m_\rho}\right)\times\frac{g^4_{F_q}}{g^4_\rho}\epsilon^4_\psi\left(\frac{m_\rho}{m_{F_q}}\right)^{2\Delta-2}.
\ea
The electroweak gauge couplings also induce non-derivative self-interactions for the Higgs, but these corrections are subleading and will be neglected.

From these results we conclude that the natural expectation is as usual $\langle H\rangle\equiv v\sim m_\rho/g_\rho$, such that a rough measure of fine-tuning in this model may be given by $g_\rho^2v^2/m_\rho^2$. The precise value of the Higgs mass, however, depends on the details of the strong dynamics.

If $\Delta^{(\alpha)}_{4\psi}\gtrsim2\Delta+2$, which is in principle possible in \emph{small} $N$ theories~\cite{Galloway:2010bp}, the IR contribution~(\ref{IR}) dominates. In this case, once the electroweak vacuum is set to its experimental value, one finds
\ba\label{higgsIR}
m_h\sim\sqrt{N_c}\frac{g_\rho}{4\pi}m_t. 
\ea
Recent LHC data~\cite{Higgs} indicate that $m_h\sim125$ GeV, and seem to require $g_\rho<4\pi$, a condition that looks slightly at odds with the assumption that the theory is at small $N$. However, no robust conclusion can be drawn without a quantitative knowledge of the numerical factors involved in these estimates. 

In large $N$ theories some 4-$\psi$ structures satisfy $\Delta^{(\alpha)}_{4\psi}\simeq2\Delta$, and the UV potential will dominate. In this case from~(\ref{UV}) we find:
\ba\label{higgsUV}\no
\sqrt{N_c}\frac{g_\rho}{4\pi}m_t\times\left(\frac{m_{F_q}}{m_\rho}\right)\lesssim\, m_h\,\lesssim\sqrt{N_c}\frac{g^2_\rho}{4\pi}v\times\frac{g_{F_q}^2}{g_\rho^2}\left(\frac{m_\rho}{m_{F_q}}\right)^{\Delta-2}.
\ea
If we require $m_{F_q}>m_\rho$ to suppress the exotic contributions to the neutron EDM, then the Higgs boson will always be heavier than in the scenarios with $m_{F_q}=m_{F_\ell}=m_\rho$. An alternative regime is defined by $m_{F_q}\sim m_\rho$ and $m_{F_\ell}\gg m_\rho$. In this case the Higgs and quark physics will resemble that of the standard scenarios, but the model will safely pass the severe bounds from the lepton sector. 

In conclusion, we find that additional, non-generic assumptions are needed to naturally accommodate a Higgs boson mass of $125$ GeV in CH models. A possibility is to abandon the overly restrictive and unrealistic assumption that these strongly coupled theories can be parametrized by a single coupling and a single mass scale, and allow the presence of light, exotic states that cutoff the Higgs mass at scales $\ll m_\rho$.

\section{5D Picture}
\label{5D}

A calculable, weakly coupled realization of our scenario can be obtained on a 5D warped background. The purpose of this short section is to present the main ingredients of this realization. 

The 5D picture is most easily understood in the limit~(\ref{epspsi}) in which Ultracolor is just the low energy remnant of the flavor dynamics. In this case the model features one approximate CFT that develops different mass scales, and may be described by a single slice of $AdS_5$ with several 4D branes. 

The $AdS_5$ space represents the Ultracolor dynamics, which enters an approximate conformal regime below the UV brane at $\Lambda_W$. The CFT is first broken by the \emph{lepton} flavor brane at the scale $m_{F_\ell}$, then by the \emph{quark} flavor brane at $m_{F_q}<m_{F_\ell}$, and finally by the \emph{Higgs} brane at $m_\rho<m_{F_q}$. The warping along the fifth dimension may differ in between the branes. The SM gauge group is a weak gauging of a subgroup of the chiral symmetry of Ultracolor, and must therefore propagate in the bulk, as usual. Similarly, the Higgs boson lives in the bulk. 

The new feature is that the quarks/leptons mix with CFT operators characterized by different compositeness scales, and hence their wavefunctions extend from the UV brane down to the corresponding quark/lepton flavor brane. 

The model reduces to the standard 5D realization of Partial Compositeness~\cite{Gherghetta:2000qt}\cite{APS} when the two flavor branes approach the IR brane at $m_\rho$, while it resembles an Extended Ultracolor scenario when the flavor branes merge with the UV. 

The stabilization of this multi-brane model may result from the introduction of bulk scalars with small 5D masses and potentials on each brane, and proceeds in a way completely analogous to the standard two-brane scenarios.

With this picture in mind the reader can qualitatively understand the results presented in Sections~\ref{general} and~\ref{realistic}, where $g_F$ plays a role analogous to the 5D Yukawa coupling and $m_F$ to the typical mass of the Kaluza-Klein fermions. In these calculable models the strongest experimental constraint comes from $K^0-\overline{K^0}$ mixing, and for $\Delta\simeq2$ it turns out to be \emph{weaker} compared to the NDA estimates~\cite{Agashe:2008uz}\cite{Gedalia:2009ws}\cite{Vecchi:2010em}.

\section{Discussion}
\label{discussion}

In this paper we have seen that natural, large $N$ Composite Higgs models benefit from the relaxation of the assumption that the mass scales characterizing the Higgs and flavor sectors coincide.


We presented scenarios in which the mass hierarchy among the three generations of quarks and leptons is explained by hierarchical mixing parameters $\epsilon_f$s, as in the conventional Partial Compositeness scenarios, while flavor violation beyond the renormalizable level is further suppressed by two new quantities: the relative mass scale $m_\rho/m_F$, and the relative strength $g_\rho/g_F$ of the Higgs and flavor sectors.

The main source of flavor violation at low energies is controlled by 4-fermion operators. Their effect can be made compatible with data if Ultracolor is in an approximately ``walking" regime in which the dimension of the ultra-fermion bilinear is $\Delta\simeq2$ and if $g_\rho^2/g_{F}^2<1$. 

The dipole and penguin interactions arise from operators with larger scaling dimensions, and are therefore suppressed by powers of $m_\rho/m_F$. This crucial fact can be employed to alleviate the constraining bounds from $\mu\to e\gamma$, the electron EDM, and the neutron EDM.

In the present framework there is no reason to expect that the lepton and quark flavor hierarchies be generated at the same scale. If the lepton flavor problem is addressed at a mass scale $m_{F_\ell}\gg m_\rho$, then lepton flavor-violating effects mediated by dipole and penguin operators decouple from the weak scale. Because the coefficients of the 4-fermion operators involving leptons are strongly suppressed by the small lepton masses, we conclude that experimentally testable signatures in the lepton sector \emph{do not} appear to be a generic implication of these models.

On the other hand, the large top mass and the irrelevant nature of the Yukawa operators severely constrain the magnitude of the quark flavor scale $m_{F_q}$, and thus effectively implies a lower bound on quark flavor violation: the existence of new sources of flavor violation in the quark sector is a robust prediction of composite Higgs models. A model-independent signature of the present framework is thus new effects in $K^0-\overline{K^0}$ mixing, while we find that $m_\rho$ of the order of a few TeV and a relatively strong flavor sector suffice for our framework to be consistent with the neutron EDM measurements.

The Higgs boson mass is highly sensitive to the details of the model. Similarly to what happens in the scenarios considered so far in the literature, non-generic conditions have to be met in order to naturally accommodate a Higgs boson mass around the value reported by the ATLAS and CMS collaborations~\cite{Higgs}.

The large $N$ nature of our framework allows us to implement the main features of this strongly coupled 4D model in a weakly coupled 5D dual theory on a slice of $AdS_5$. Besides the usual UV and an IR branes, the present model develops intermediated flavor branes at $m_{F_q}\gtrsim m_\rho$ and $m_{F_\ell}\gg m_{F_q}$.


\begin{acknowledgments}
I thank Silvia for her patience, Marco Nardecchia and Ian Shoemaker for discussions, and Kaustubh Agashe for comments, for suggesting a brief discussion of the 5D realization, and for helping me in the identification of the dual picture. This work was supported by the DOE Office of Science and the LANL LDRD program. 
\end{acknowledgments}


\end{document}